# Universal Framework for Quantum Error-Correcting Codes

Zhuo Li and Li-Juan Xing

State Key Laboratory of Integrated Service Networks, Xidian University, Xi'an,

Shannxi 710071, China

We present a universal framework for quantum error-correcting codes, i.e., the one that applies for the most general quantum error-correcting codes. This framework is established on the group algebra, an algebraic notation for the nice error bases of quantum systems. The nicest thing about this framework is that we can characterize the properties of quantum codes by the properties of the group algebra. We show how it characterizes the properties of quantum codes as well as generates some new results about quantum codes.

PACS numbers: 03.67.Pp, 03.67.Hk, 03.67.Lx

## I. INTRODUCTION

The quest to build a scalable quantum computer that is resilient against decoherence errors and operational noise has sparked a lot of interest in quantum error-correcting codes [1-9]. Although arbitrary error operators might affect a quantum state, it is always possible to keep track of the error amplitudes by expressing them in terms of an error operator basis. A particularly useful class of unitary error bases, called nice error bases, has been introduced by Knill in [10]. The nice error bases are the pillar of quantum error-correcting codes [11].

When a new physical problem occurs, it is always desirable to find an appropriate framework for it, such as quantum mechanics for quantum physics. Since the occurrence of quantum codes, almost all the researches are carried out on the specific types of quantum codes, for example, mainly on stabilizer codes, pure codes and codes over finite field. In this paper we are mainly interested in universal framework for quantum codes, i.e., the one which applies for all codes, no matter they are pure or not, stabilizer codes or not, over finite field or not. Firstly we recall the properties of nice error bases. Then we give the definitions of the group algebra and characters associated with nice error basis. Finally, based on the group algebra, we establish a universal framework for quantum codes. Through the discussion we show this framework can characterizes the properties of quantum codes as well as generates some new results about quantum codes. It is a powerful tool in future works on quantum codes.

## II. PRELIMINARIES

Quantum information can be protected by encoding it into a quantum error-correcting code. An $((n,K,d))_m$ quantum code is a $K$-dimensional subspace of the state space of $n$ quantum systems with $m$ levels that can detect all errors affecting less than $d$ quantum systems, but cannot detect some errors affecting $d$ quantum systems.

Let $\mathcal{H} = \mathbb{C}^m$ be a quantum system with $m$ levels and let $G$ be an additive group of order $m^2$ with identity element 0. A nice error basis of $\mathcal{H}$ is a set $\mathcal{E} = \{E_g \mid g \in G\}$ of unitary operators on $\mathcal{H}$ such that

i) $E_0$ is the identity operator,

ii) $\text{tr}\, E_g = n\delta_{g,0}$ for all $g \in G$,

iii) $E_g E_h = \omega_{gh} E_{g+h}$ for all $g, h \in G$,

where complex numbers $\omega_{gh}$ have modulus 1. We call $G$ the index group of the error basis $\mathcal{E}$. Moreover $\mathcal{E}_n \triangleq \mathcal{E}^{\otimes n} = \{E_g \triangleq E_{g_1} \otimes \cdots \otimes E_{g_n} \mid g = (g_1, \ldots, g_n) \in G^n\}$ is a nice error basis of $n$ quantum systems $\mathcal{H}^{\otimes n}$.

***Lemma 1.*** If the index group $G$ is Abelian, then for any nonzero $h \in G$,
$$\sum_{g \in G} \omega_{gh} \overline{\omega}_{hg} = 0.$$

***Proof.*** From property iii) of the nice error basis, it follows that
$$E_a E_b E_h = (\omega_{bh} \overline{\omega}_{hb}) E_a E_h E_b = (\omega_{ah} \overline{\omega}_{ha})(\omega_{bh} \overline{\omega}_{hb}) E_h E_a E_b$$

and
$$E_a E_b E_h = \omega_{ab} E_{a+b} E_h = \omega_{ab}(\omega_{(a+b)h} \overline{\omega}_{h(a+b)}) E_h E_{a+b} = (\omega_{(a+b)h} \overline{\omega}_{h(a+b)}) E_h E_a E_b,$$

which means
$$(\omega_{ah} \overline{\omega}_{ha})(\omega_{bh} \overline{\omega}_{hb}) = \omega_{(a+b)h} \overline{\omega}_{h(a+b)}. \tag{1}$$

Now let $G_h = \{\omega_{gh} \overline{\omega}_{hg} \mid g \in G\}$. Then from (1) $G_h$ is a subgroup of a cyclic group since all $\omega_{gh}$ generate a cyclic group [12]. Thus $G_h$ itself is a nontrivial cyclic group for any nonzero $h \in G$.  Q.E.D.

In the next section we shall give the concept of the group algebra based on the nice error basis with the Abelian index group. For simplicity, we assume throughout the paper that the index group $G$ of the error basis $\mathcal{E}$ is Abelian. This assumption is reasonable because such nice error basis exists for a quantum system with arbitrary $m$ levels [10].

## III. GROUP ALGEBRA

We are going to describe the elements of $\mathcal{E}_n$ by formal polynomials in $z_1, \ldots, z_n$. In general $E_g = E_{g_1} \otimes \cdots \otimes E_{g_n}$ is represented by $z_1^{g_1} z_2^{g_2} \cdots z_n^{g_n}$, which we abbreviate $z^g$. We make the convention that $z_i^{g_i} z_i^{h_i} = z_i^{g_i + h_i}$. This makes the set of all $z^g$ into a multiplicative group denoted by $Z$. Thus $G^n$ and $Z$ are isomorphic groups, with addition in $G^n$

$$g + h = (g_1, \ldots, g_n) + (h_1, \ldots, h_n) = (g_1 + h_1, \ldots, g_n + h_n)$$

corresponding to multiplication in $Z$

$$z^g z^h = z_1^{g_1} \cdots z_n^{g_n} \cdot z_1^{h_1} \cdots z_n^{h_n} = z_1^{g_1 + h_1} \cdots z_n^{g_n + h_n} = z^{g+h}.$$

***Definition 2.*** The group algebra $\mathbb{C}Z$ of $Z$ over the complex numbers $\mathbb{C}$ consists of all formal sums

$$\sum_{g \in G^n} a_g z^g, \quad a_g \in \mathbb{C}, \quad z^g \in Z.$$

Addition and multiplication of elements of $\mathbb{C}Z$ are defined in the natural way by

$$\sum_{g \in G^n} a_g z^g + \sum_{g \in G^n} b_g z^g = \sum_{g \in G^n} (a_g + b_g) z^g,$$

$$r \sum_{g \in G^n} a_g z^g = \sum_{g \in G^n} r a_g z^g, \quad r \in \mathbb{C}$$

and

$$\sum_{g \in G^n} a_g z^g \cdot \sum_{h \in G^n} b_h z^h = \sum_{g, h \in G^n} a_g b_h z^{g+h}.$$

To each $h \in G^n$ we associate the mapping $\chi_h$ from $Z$ to the complex numbers given by

$$\chi_h(z^g) = \operatorname{tr} E_h^\dagger E_g^\dagger E_h E_g / m^n,$$

$\chi_h$ is called a character of $Z$. $\chi_h$ is extended to act on $\mathbb{C}Z$ by linearity:

$$\chi_h\left(\sum_{g \in G^n} a_g z^g\right) = \sum_{g \in G^n} a_g \chi_h(z^g) = \sum_{g \in G^n} a_g \operatorname{tr} E_h^\dagger E_g^\dagger E_h E_g / m^n.$$

Note that

$$\chi_h(z^g) = \prod_{i=1}^{n} \omega_{h_i g_i} \bar{\omega}_{g_i h_i} . \qquad (2)$$

Let

$$C = \sum_{g \in G^n} c_g z^g$$

be an arbitrary element of the group algebra $\mathbb{C}Z$, with the property that

$$M = \sum_{g \in G^n} c_g \neq 0 .$$

***Definition 3.*** The transform of $C$ is the element $C'$ of $\mathbb{C}Z$ given by

$$C' = \frac{1}{M} \sum_{h \in G^n} \chi_h(C) z^h ,$$

where $\chi$ was defined above.

Suppose

$$C' = \sum_{h \in G^n} c'_h z^h ,$$

then

$$c'_h = \frac{1}{M} \chi_h(C) = \frac{1}{M} \sum_{g \in G^n} c_g \operatorname{tr} E_h^\dagger E_g^\dagger E_h E_g / m^n , \quad h \in G^n , \qquad (3)$$

and $c'_0 = 1$. Now we describe several weight enumerators of the group algebra $\mathbb{C}Z$.

Let the elements of $G$ be denoted by $\alpha_0 = 0, \alpha_1, \ldots, \alpha_{m^2-1}$, in some fixed order.

The first weight enumerator to be considered specifies the group algebra completely by introducing enough variables. In general, the variables $z_{ij}$ means that the $i^{\text{th}}$ place in the vector $g$ is the $j^{\text{th}}$ element $\alpha_j$ of $G$. The vector $g = (\alpha_{a_1}, \alpha_{a_2}, \ldots, \alpha_{a_n})$ is described by the polynomial

$$f(g) = z_{1 a_1} z_{2 a_2} \cdots z_{n a_n} .$$

Thus $g$ is uniquely determined by $f(g)$. This requires the use of $nm^2$ variables $z_{ij}$, $1 \le i \le n, 0 \le j \le m^2 - 1$.

What we shall call the exact enumerator of $C$ is then defined as

$$\mathcal{E}_C = \sum_{g \in G^n} c_g f(g).$$

Then the exact enumerator of $C'$ is

$$\mathcal{E}_{C'} = \sum_{h \in G^n} c'_h f(h).$$

***Theorem 4.***

$$\mathcal{E}_{C'}(z_{10}, \ldots, z_{ir}, \ldots, z_{n(m^2-1)}) = \frac{1}{M} \mathcal{E}_C \left( \sum_{s=0}^{m^2-1} \omega_{\alpha_s \alpha_0} \overline{\omega}_{\alpha_0 \alpha_s} z_{1s}, \ldots, \sum_{s=0}^{m^2-1} \omega_{\alpha_s \alpha_r} \overline{\omega}_{\alpha_r \alpha_s} z_{is}, \ldots, \sum_{s=0}^{m^2-1} \omega_{\alpha_s \alpha_{m^2-1}} \overline{\omega}_{\alpha_{m^2-1} \alpha_s} z_{ns} \right).$$

***Proof.*** From (2) and (3), the LHS is equal to

$$\sum_{h \in G^n} c'_h f(h) = \frac{1}{M} \sum_{g \in G^n} c_g \sum_{h \in G^n} \chi_h(z^g) f(h) = \frac{1}{M} \sum_{g \in G^n} c_g \sum_{s_1=0}^{m^2-1} \sum_{s_2=0}^{m^2-1} \cdots \sum_{s_n=0}^{m^2-1} \prod_{i=1}^{n} \omega_{\alpha_{s_i} g_i} \overline{\omega}_{g_i \alpha_{s_i}} z_{is_i}$$

$$= \frac{1}{M} \sum_{g \in G^n} c_g \prod_{i=1}^{n} \sum_{s=0}^{m^2-1} \omega_{\alpha_s g_i} \overline{\omega}_{g_i \alpha_s} z_{is}$$

which is equal to the RHS.  Q.E.D.

The next weight enumerator to be considered classifies vectors $g$ in $G^n$ according to the number of times each group element $\alpha_i$ appears in $g$.

***Definition 5.*** The composition of $g = (g_1, \ldots, g_n)$, denoted by $\text{comp}(g)$, is $(s_0, s_1, \ldots, s_{m^2-1})$ where $s_i = s_i(g)$ is the number of components $g_j$ equal to $\alpha_i$. Clearly

$$\sum_{i=0}^{m^2-1} s_i = n.$$

We call the set $\{A(t)\}$ the complete weight distribution of $C$ where $A(t)$ is the sum of $c_g$ with $\text{comp}(g) = t = (t_0, \ldots, t_{m^2-1})$. We also define the complete weight enumerator of $C$ to be

$$\mathcal{W}_C(z_0, \ldots, z_{m^2-1}) = \sum_t A(t) z_0^{t_0} \cdots z_{m^2-1}^{t_{m^2-1}} = \sum_{g \in G^n} c_g z_0^{s_0} \cdots z_{m^2-1}^{s_{m^2-1}}.$$

Then the complete weight distribution of $C'$ is $\{A'(t)\}$, where $A'(t)$ is the sum of $c'_h$ with $\text{comp}(h) = t = (t_0, \ldots, t_{m^2-1})$, and the complete weight enumerator of $C'$ is

$$\mathcal{W}_{C'}(z_0, \ldots, z_{m^2-1}) = \sum_t A'(t) z_0^{t_0} \cdots z_{m^2-1}^{t_{m^2-1}}.$$

**Theorem 6.**

$$\mathcal{W}_{C'}(z_0, \ldots, z_r, \ldots, z_{m^2-1}) = \frac{1}{M} \mathcal{W}_C\left( \sum_{s=0}^{m^2-1} \omega_{\alpha_s \alpha_0} \overline{\omega}_{\alpha_0 \alpha_s} z_s, \ldots, \sum_{s=0}^{m^2-1} \omega_{\alpha_s \alpha_r} \overline{\omega}_{\alpha_r \alpha_s} z_s, \ldots, \sum_{s=0}^{m^2-1} \omega_{\alpha_s \alpha_{m^2-1}} \overline{\omega}_{\alpha_{m^2-1} \alpha_s} z_s \right).$$

*Proof.* Set $z_{ij} = z_j$ for $1 \le i \le n, 0 \le j \le m^2 - 1$ in Theorem 4. Q.E.D.

By setting certain variables equal to each other in the complete weight enumerator we obtain the Lee and Hamming weight enumerators, which give progressively less and less information about the group algebra, but become easier to handle.

**Definition 7.** Suppose now that $m^2 = 2\delta + 1$ is odd, and let the elements of $G$ be labeled $\alpha_0 = 0, \alpha_1, \ldots, \alpha_\delta, \alpha_{\delta+1}, \ldots, \alpha_{m^2-1}$, where $\alpha_{m^2-i} = -\alpha_i$ for $1 \le i \le \delta$. The Lee composition of a vector $g \in G^n$, denoted by $\text{Lee}(g)$, is $(l_0, l_1, \ldots, l_\delta)$ where $l_0 = s_0(g)$, $l_i = s_i(g) + s_{m^2-i}(g)$ for $1 \le i \le \delta$.

We call the set $\{L(t)\}$ the Lee weight distribution of $C$ where $L(t)$ is the sum of $c_g$ with $\text{Lee}(g) = t = (t_0, \ldots, t_\delta)$. We also define the Lee weight enumerator of $C$ to be

$$\mathcal{L}_C(z_0, \ldots, z_\delta) = \sum_t L(t) z_0^{t_0} z_1^{t_1} \cdots z_\delta^{t_\delta} = \sum_{g \in G^n} c_g z_0^{l_0} z_1^{l_1} \cdots z_\delta^{l_\delta}.$$

Then the Lee weight distribution of $C'$ is $\{L'(t)\}$, where $L'(t)$ is the sum of $c'_h$ with $\text{Lee}(h) = t = (t_0, \ldots, t_\delta)$, and the Lee weight enumerator of $C'$ is

$$\mathcal{L}_{C'}(z_0, \ldots, z_\delta) = \sum_t L'(t) z_0^{t_0} z_1^{t_1} \cdots z_\delta^{t_\delta}.$$

***Theorem 8.*** The Lee enumerator for the transform $C'$ is obtained from the Lee enumerator of $C$ by replacing each $z_i$ by

$$z_0 + \sum_{s=1}^{\delta} (\omega_{\alpha_s \alpha_i} \bar{\omega}_{\alpha_i \alpha_s} + \bar{\omega}_{\alpha_s \alpha_i} \omega_{\alpha_i \alpha_s}) z_s,$$

and dividing the result by $M$.

***Proof.*** Set $z_{m^2 - i} = z_i$ for $1 \leq i \leq \delta$ in Theorem 6.  Q.E.D.

The Hamming weight, or simply the weight, of a vector $g = (g_1, \ldots, g_n) \in G^n$ is the number of nonzero components $g_i$, and is denoted by $\text{wt}(g)$.

We call the set $\{A_i\}$ the Hamming weight distribution of $C$ where $A_i$ is the sum of $c_g$ with $\text{wt}(g) = i$. We also define the Hamming weight enumerator of $C$ to be

$$W_C(x, y) = \sum_{i=0}^{n} A_i x^{n-i} y^i = \sum_{g \in G^n} c_g x^{n - \text{wt}(g)} y^{\text{wt}(g)}.$$

Then the Hamming weight distribution of $C'$ is $\{A'_i\}$, where $A'_i$ is the sum of $c'_h$ with $\text{wt}(h) = i$, and the Hamming weight enumerator of $C'$ is

$$W_{C'}(x, y) = \sum_{i=0}^{n} A'_i x^{n-i} y^i.$$

***Theorem 9.***

$$W_{C'}(x, y) = \frac{1}{M} W_C(x + (m^2 - 1)y, x - y).$$

***Proof.*** In Theorem 6 put $z_0 = x$, $z_1 = z_2 = \cdots = z_{m^2 - 1} = y$, and use Lemma 1.  Q.E.D.

# IV. UNIVERSAL FRAMEWORK FOR QUANTUM CODES

In this section, we establish the universal framework for quantum codes based on the group algebra defined in the last section.

Given an arbitrary quantum code $\mathcal{C} = ((n, K, d))_m$, let $P = \sum_{i=1}^{K} |v_i\rangle\langle v_i|$ be the orthogonal projection onto $\mathcal{C}$ where $\{v_i\}$ is a set of orthonormal basis of $\mathcal{C}$, and let $G$ be the index group of any nice error basis $\mathcal{E}$ of the quantum system with $m$ levels. Then we can formulize the quantum code $\mathcal{C}$ as an element $C = \sum_{g \in G^n} c_g z^g$ from the group algebra $\mathbb{C}Z$ where

$$c_g = \frac{1}{K^2}(\operatorname{tr} E_g P^\dagger)(\operatorname{tr} E_g^\dagger P) = \frac{1}{K^2} \left| \sum_{i=1}^{K} \langle v_i | E_g | v_i \rangle \right|^2. \tag{4}$$

We call $C$ the element associated with the quantum code $\mathcal{C}$ in the group algebra $\mathbb{C}Z$.

From (3), the transform of $C$ is given by $C' = \sum_{h \in G^n} c'_h z^h$ where

$$c'_h = \frac{1}{M} \sum_{g \in G^n} \frac{1}{K^2} (\operatorname{tr} E_g P^\dagger)(\operatorname{tr} E_g^\dagger P) \operatorname{tr} E_h^\dagger E_g^\dagger E_h E_g \Big/ m^n$$

$$= \frac{m^n}{K^2 M} \operatorname{tr} E_h^\dagger \mathcal{P}^\dagger E_h \mathcal{P} = \frac{m^n}{K^2 M} \sum_{i=1}^{K} \sum_{j=1}^{K} |\langle v_i | E_h | v_j \rangle|^2 \tag{5}$$

where $M = \sum_{g \in G^n} c_g$. Since $c'_0 = 1$, from (5) we get $M = m^n / K$. Thus

$$c'_h = \frac{1}{K} \sum_{i=1}^{K} \sum_{j=1}^{K} |\langle v_i | E_h | v_j \rangle|^2. \tag{6}$$

From (4), (6), and using the Cauchy-Schwartz inequality we deduce that $c_g \leq c'_g$ for all $g \in G^n$. Furthermore, from the definition of the minimum distance $d$ we get that if $K > 1$ then $c_g = c'_g$ for all $g$ satisfying $\operatorname{wt}(g) < d$ and there exist some $g$ with $\text{wt}(g) = d$ such that $c_g \neq c'_g$; if $K = 1$ then $c_g = c'_g$ for all $g \in G^n$ and the minimum nonzero weight of $g$ such that $c_g \neq 0$ is $d$.

So far we have established the universal framework for quantum codes:

For arbitrary quantum code $\mathcal{C} = ((n, K, d))_m$ we can characterize it as the element $C = \sum_{g \in G^n} c_g z^g$ of the group algebra $\mathbb{C}Z$, called the element associated with the quantum code $\mathcal{C}$, and the transform $C' = \sum_{h \in G^n} c'_h z^h$ of $C$ so that

1) the dimension $K$ of $\mathcal{C}$ equals $m^n/M$ where $M = \sum_{g \in G^n} c_g$,

2) the minimum distance $d$ of $\mathcal{C}$ equals the minimum weight of $g$ such that $c_g \neq c'_g$ if $K > 1$; the minimum nonzero weight of $g$ such that $c_g \neq 0$ if $K = 1$.

## V. DISCUSSION

The nicest thing about the framework is that we can characterize the properties of quantum codes by the properties of the group algebra. So the problems about unfamiliar quantum codes can be transformed into those about familiar classical group algebra.

For example, we can define the weight distributions of the quantum code $\mathcal{C}$ as the weight distributions of the element $C$ associated with $\mathcal{C}$ in the group algebra $\mathbb{C}Z$ and define the dual weight distributions of $\mathcal{C}$ as the weight distributions of the transform $C'$ of $C$. Then for any quantum code, its weight distributions and dual weight distributions must satisfy the identities in Theorem 4, 6, 8, and 9. Note that the results about exact enumerators, complete enumerators and Lee enumerators of quantum codes are completely new. For Hamming weight enumerators, the binary version was first proved for quantum stabilizer codes by Calderbank et al. in [13], and later generalized by Rains in [14]. The nonbinary version for stabilizer codes was proved by Ketkar et al. in [15]. The result given here is a generalization to the most general quantum codes.

Again the purity of quantum codes can also be characterized by the group algebra. For arbitrary quantum code $\mathcal{C} = ((n, K, d))_m$, let $C = \sum_{g \in G^n} c_g z^g$ be the element associated with $\mathcal{C}$ in the group algebra $\mathbb{C}Z$. Then $\mathcal{C}$ is pure if and only if $c_g = 0$ for $0 < \text{wt}(g) < d$.

Finally if $\mathcal{C}$ is a quantum stabilizer code, from the definition of stabilizer codes, the element associated with $\mathcal{C}$ in the group algebra $\mathbb{C}Z$ can be written as $C = \sum_g z^g$ where the summation is over all such $g$ that the operator $E_g$ belongs to the stabilizer of $\mathcal{C}$. In addition, the transform of $C$ can be written as $C' = \sum_h z^h$ where the summation is over all such $h$ that the operator $E_h$ belongs to the normalizer of $\mathcal{C}$. Both forms imply the relationship between quantum stabilizer codes and classical codes.

To sum up, we have presented a universal framework for quantum codes and shown how it characterizes the properties of quantum codes as well as generates new results about quantum codes. We can assert that this framework is a very useful and potential tool in studying the problems about quantum error-correcting codes.